\documentclass[11pt]{article}
\usepackage{setspace}
\usepackage{amssymb}
\usepackage{amsmath}
\usepackage{amsfonts}

\def\be{\begin{equation}}

\def\ee{\end{equation}}
\newcommand\prefive{\, ^5}

\def\dd{\partial}

\newcommand\of[1]{\left( #1 \right)}
\newcommand\sqof[1]{\left[ #1 \right]}
\newcommand\E{{\mathcal G}}
\newcommand\R{{\mathcal R}}

\def\bea{\begin{eqnarray}}

\def\eea{\end{eqnarray}}

\newcommand\eps{\epsilon}

\begin{document}

\singlespace

\begin{flushright} BRX TH-640 \\
CALT 68-2851
\end{flushright}

\vspace*{.3in}

\begin{center}

{\Large\bf  Canonical Analysis and Stability of Lanczos-Lovelock Gravity}

{\large S.\ Deser}

{\it Physics Department,  Brandeis University, Waltham, MA 02454 and \\
Lauritsen Laboratory, California Institute of Technology, Pasadena, CA 91125 \\
{\tt deser@brandeis.edu}
}

{\large J.\ Franklin}

{\it  Reed College, Portland, OR 97202 \\
{\tt jfrankli@reed.edu}}

\end{center}

\begin{abstract}
We perform a space-time analysis of the $D>4$ quadratic curvature Lanczos-Lovelock (LL) model, exhibiting its dependence on intrinsic/extrinsic curvatures, lapse and shifts.  As expected from general covariance, the field equations include $D$ constraints, of zeroth and first time derivative order. In the ``linearized" -- here necessarily cubic -- limit, we give an explicit formulation in terms of the usual ADM  metric decomposition, incidentally showing that time derivatives act only on its transverse-traceless spatial components. Unsurprisingly, pure LL has no Hamiltonian formulation, nor are even its  -- quadratic -- weak field constraints easily soluble. 
Separately, we point out that the extended, more physical $R+$LL, model is stable -- its energy is positive -- due to its supersymmetric origin and ghost-freedom.

\end{abstract}

\section{Introduction}

The, necessarily $D>4$,  Lanczos-Lovelock (LL) quadratic curvature model [1] is unique among quadratic curvature theories in being ghost-free and of second derivative order, as well as representing the first string correction to the Einstein action [2]. It has since been primarily studied, following [3], in the context of special solutions. Here, we consider the theory's complementary, dynamical, properties, by making a ``$(D-1)+1$" ADM [4] decomposition of the pure LL action (for simplicity) in terms of intrinsic and extrinsic (second fundamental form) curvatures, lapse and shifts. The theory's intrinsic nonlinearity -- even its lowest order action is necessarily cubic rather than quadratic -- is an immediate sign of its exotic character. Nevertheless, understanding its dependence on the embedding components is instructive: for example, it exhibits universal aspects of coordinate invariant theories, including the presence of $D$ lower time derivative order constraints, i.e., lower derivative field equations;  we write these explicitly.  What the LL action -- unsurprisingly -- lacks is a Hamiltonian form. We will also analyze its ``linearized" -- cubic -- limit in terms of the ADM ``TT" metric decomposition, which reveals that only the $g^{TT}_{ij}$ components carry time derivatives.  While the action and field equations can be written quite explicitly in terms of the metric, its cubic kinetic terms and quadratic constraints differentiate it from normal models.  The ``TT" decomposition will also be used to provide an alternative understanding of ``Birkhoff's" theorem [6] here, and for the full theory.

Separately, we consider the extended, $R+$LL, theory.  We point out that its supersymmetrizability -- guaranteed by its superstring origin -- plus manifest ghost-freedom imply positive energy, just as in GR [9], for the ``branch" with proper relative LL to $R$ actions' sign.  Some concrete properties of the underlying energy constraint are also given.

\section{LL and its space-time decomposition}

The  LL action's normal covariant form is
\begin{equation}
I[g_{\mu\nu}] = \int d^5x \,  \eps^{\alpha \beta \delta \gamma \mu} \, \eps^{\kappa \rho \sigma \tau \nu} \, R_{\alpha\beta\kappa\rho} \, R_{\delta\gamma\sigma\tau} \, g_{\mu\nu}/\sqrt{-g}.
\end{equation}

Our conventions are: signature mostly plus, greek/roman indices cover $5/4$ dimensions,
$R^\mu_{\nu\alpha\beta}\sim + \dd_\alpha \, \Gamma^\mu_{\nu\beta} + \ldots$, the contravariant Levi-Civita density $\eps^{01234} =+1$; we have set the overall, dimensional, gravitational constant to unity. For concreteness (only), we work in $D=5$, the lowest non-trivial dimension.  [One could also proceed in terms of the vielbein form,
\begin{equation}
\begin{aligned} 
I[ e_{\nu \alpha}] &= \int d^5x \,  \eps^{abcde} \, \eps^{\mu\nu\rho \sigma \tau} \, R_{\mu\nu ab} \, R_{\rho\sigma cd} \, e_{\tau e}, \\
R_{\mu \nu ab} &\equiv \of{ D_\mu \, \omega_{\nu ab} +  \omega^{\, \, \, c}_{\mu \, \, \, \, a} \, \omega_{\nu c b}  + \hbox{symm.} }, \, \, \, \, \, \, \, \, \, D(w)_\nu e_{\mu a} &\equiv 0;
\end{aligned}
\end{equation}
here roman indices are local; however, as in GR [7], this formalism is no simpler than the metric one.]

The special virtue of LL is that it is the unique quadratic action whose field equations do not involve derivatives of the curvature, hence contain no higher than second derivatives of the metric. [As a reminder, the $R$-variations of (1) do not contribute, because 
\begin{equation}
\delta R_{\alpha\beta\gamma\delta} =1/2 \,  (D_\gamma D_\beta \, \delta g_{\alpha \delta} + \hbox{symm});
\end{equation}
these derivatives contribute neither when they land on the other $R$ -- by the uncontracted Bianchi identities -- nor when acting on the undifferentiated metric, since $D_\gamma \, g_{\mu\nu} \equiv 0$.]  Varying the remaining variable, $g_{\mu\nu}$, in (1) then trivially yields the field equations\footnote{As in any other theory, choice of basic fields -- here metric components -- and index positions would lead to a reshuffling of the field equations, but not their overall contents.}
\begin{equation}
\E^{\mu\nu} \equiv \, \of{ \R^{\mu\nu} - 1/2 \, g^{\mu\nu} \, \R^\gamma_\gamma} = 0, \, \, \, \, \, \, \, \, \, 
\R^{\mu\nu} \equiv  \eps^{\alpha \beta \delta \gamma \mu} \, \eps^{\kappa \rho \sigma \tau \nu} \, \R_{\alpha\beta\kappa\rho} \, \R_{\delta\gamma\sigma\tau}/(-g).
\end{equation}
We have written (4) to parallel GR: The ``Einstein tensor's" first term is the ``Ricci tensor". Note that the field equations obey the Bianchi-like identity, $D_\mu \, \E^{\mu\nu} \equiv 0$ -- simply by the action's manifest coordinate invariance -- thus ensuring, just as in GR, that the ($\E^{00}$/$\E^{0i}$) are of lower -- ($0$/$1$) -- time derivative order than the second-order $\E^{ij}$, as we shall see explicitly below. [As a reminder, in Riemann normal coordinates, $\dd_0 \, \E^{0i} = -\dd_j \, \E^{ij}$, so $\E^{0i}$ is one time derivative order lower than $\E^{ij}$, while $\dd_0 \, \E^{00} = -\dd_i \, \E^{i 0}$ puts $\E^{00}$ one order below $\E^{0i}$.]

Next, we turn to the promised space-time decomposition.  That of the metric is the usual ADM one,
\begin{equation}
\prefive g_{\mu\nu} = [ g_{ij}, g_{0i} \equiv N_i, g_{00} \equiv -(N^2 - N^s \, N_s)],  \, \, \, \, \, \, \, \, \, \sqrt{-\prefive g} = N \, \sqrt{g}.
\end{equation}
In terms of these variables, the standard Gauss-Codazzi embedding apparatus (see, e.g., [8]) then expresses the space-time curvature components $\prefive R_{\mu\nu\alpha\beta}$ in terms of their two building blocks: intrinsic $D=4$ spatial curvature $R_{ijkl}(g_{pq})$ and second fundamental form/extrinsic curvature $K_{ij}$,
\begin{equation}
K_{ij} \equiv- 1/(2 \, N) \, \of{ \dot g_{ij} - N_{i|j} - N_{j|i}};
\end{equation}
of these, only $K_{ij}$ carries time derivatives.   Specifically, 
\begin{eqnarray}
\prefive R_{ijkl} &=& R_{ijkl} - \of{ K_{il} \, K_{jk} - K_{ik} \, K_{jl} } \\
\prefive R_{0ijk} &=& \sqof{ R_{jksi}  - \of{K_{ij} \, K_{sk} - K_{ik}\, K_{js}}}\, N^s - \sqof{K_{ji|k} - K_{ki|j}} \, N \\
\prefive R_{0i0j} &=& \sqof{ R_{jpis}  - \of{K_{is} \, K_{jp} -K_{ij}\, K_{sp}} } \, N^s \, N^p - \sqof{ K_{si|j} + K_{sj|i} - 2 \, K_{ij|s} } \, N\, N^s \\
 &+& \sqof{\dot K_{ij}+ K^s_i \, K_{js}} \, N^2 + N \, N_{|ij}. \nonumber
\end{eqnarray}
The resulting space-time form of (1,4) is obtained upon inserting the above ``$4+1$" decomposition there.  The time derivatives, as tracked by 
$K_{ij}$ and $\dot K_{ij}$, verify the lower order of the $\R^{0\mu}$ equations, as predicted.  Also, as mentioned earlier, LL  has no Hamiltonian, ``$p\,  \dot q- N\, R$", form: even the linearized limit is cubic, since -- by the discussion between (3) and (4) above -- the linearized Lagrangian, 
$L_{\hbox{\tiny{lin}}} \equiv \eps\,  \eps \, R^L \, R^L \, \eta_{\mu\nu}$, is a total divergence in any $D$.

 In the weak field limit, to which we turn next, we will be able to go beyond the above curvature description to a concrete formulation in terms of the metric deviation and its ``TT" decomposition.
\section{Weak field limit}

In the weak field limit, where we keep only lowest order in the metric deviation $h_{\mu\nu} \equiv g_{\mu\nu} - \eta_{\mu\nu}$, the curvature
linearizes to 
\begin{equation}
\prefive R_{\alpha \beta \gamma \delta} = 1/2 \, \sqof{ h_{\alpha \delta, \beta \gamma} - h_{\delta \beta, \alpha \gamma} - h_{\alpha \gamma, \beta \delta} + h_{\gamma \beta, \alpha \delta}},
\end{equation}
and the action reduces to the schematic form
\begin{equation}
I \sim \int d^5 x \, \eps \, \eps \, R^L \, R^L h \sim  \int d^5x\, \eps\,  \eps\,  \dd \dd  h\, \dd \dd h\,  h.
\end{equation}
[To make this fact quite clear, recall that the full action varies into the full $R\, R=0$ equations, and that this holds order by order in $h_{\mu\nu}$. Hence to lowest, quadratic, order the field equations read $\eps\,  \eps\,  R^L\,  R^L= 0$ and come from the combined variation of all the action's cubic terms, $I_C \sim \int \eps\,  \eps \of{R^L \, R^L\,  h +R^Q \, R^L\,  \eta}$.].  Using (10), the components (7-9) can be given explicitly:
\begin{eqnarray}
\prefive R_{ijkl} &=& R_{ijkl}(h_{mn}) =1/2 \, [ h_{il,jk} + \hbox{symm}], \, \, \, \, \, \, \, \, \,  \prefive R_{0ijk} =(K_{ki,j} - K_{ji,k}), \, \, \, \, \, \, \, \, \, \,  \prefive R_{0i0j} = \of{ \dot K_{ij} + n_{,ij} }, \nonumber \\
 K_{ij} &=& -1/2\sqof{ \dot h_{ij} -N_{i,j}-N_{j,i}}, \, \, \, \, \, \, \, \, \, n = N-1 = -1/2 \, h_{00}, \, \, \, \, \, \, \, \, \, \sqrt{-g} = 1 + 1/2\, (h^T + 2 \, n).
\end{eqnarray}

We also perform the usual ``TT" decomposition of the metric deviation:
\begin{equation}
h_{ij} = h^{TT}_{ij} + h^T_{ij} +(h_{i,j}+h_{j,i}), \, \, \, \,     h^{TT}_{ij,j} \equiv 0 \equiv h^{TT}_ {ii}, 
\, \, \, \,  h^T_{ij} \equiv 1/3 \, (\delta_{ij}- \dd_i \dd_j/\nabla^2)\, h^T, \, \, \, \, 
h_{0i} \equiv N^T_i +N^L_{,i}.
\end{equation}
Since the action (11) is invariant under the linearized coordinate transformation, $h_{\mu\nu}\longrightarrow X_{\mu,\nu} +X_{\nu,\mu}$, the $h_i$ component in (13) is absent, and only $(h^{TT}_{ij}, h^T, n, N_i)$ appear. The remaining gauge freedom may be used, as in GR, to choose the time gauge so that $K^T=0$, leaving 
\begin{equation}
K_{ij} = -1/2 \sqof{ \dot h^{TT}_{ij} -N_{i,j} - N_{j,i}}. 
\end{equation}
Thus, $h^{TT}$ is the only weak field metric component carrying time derivatives. This immediately links Birkhoff's theorem [6] to the identical vanishing of spherically symmetric ``TT"-tensors, since only the latter carry time derivatives.  Indeed, we can 
use the same argument for the complete model: Given its coordinate-invariance, we may set the time 
gauge choice so as to keep the full $K_{ij}$ free of any but the $g_{ij}$'s ``TT" component; then the -- purely 
kinematical -- absence of spherically symmetric ``TT" tensors ensures the Birkhoff result.  Just as in GR, 
there is no obstacle to making the ``TT" decomposition at full nonlinear level.

A last step to expressing the action in terms of the metric deviation is to note that $R_{ijkl}$ depends only on the combination $h^{\hbox{\tiny{eff}}}_{ij} \equiv h^{TT}_{ij} + h_{ij}^T$ since linearized curvature is just the double curl of $h_{ij}$.
With this machinery, we can at last express the linearized action (11) explicitly,
\begin{equation}
\begin{aligned}
I&[ h^{TT}, h^T, n,N_i]= 144\,  \int d^5x\,   \eps^{0ijkl} \, \eps^{0pqrs}  \, \biggl[  \of{h_{l s} - 1/2 \, \delta_{l s} (h^T + 2 \, n)}  \, \bigl\{  \dot h^{TT}_{r i, q} \, \dot h^{TT}_{kp,j} - 2 \,N_{k, p j}\,  ( \dot h^{TT}_{r i, q } - N_{r, i q} )  \\
& + ( -\ddot h^{TT}_{k r} +  2 \, \dot N_{k,r}  + 2\, n_{,k r} ) \, h_{iq, j p}\bigr\} - 1/2 \, (n- 1/2 \, h^T) \, h_{i s, l r} \, h_{j q, k p} - 2 \, N_l \, \of{ \dot h^{TT}_{s k, r} - N_{s, k r} } \, h_{i q, j p}  \biggr].
 \end{aligned}
\end{equation}
The resulting field equations are, in terms of $\R^{\mu\nu}$ for convenience,
\begin{eqnarray}
\R^{l s} &=& \eps \, \eps \,  \biggl[ \dot h^{TT}_{r i, q} \, \dot h^{TT}_{kp,j} -2 \, N^T_{k,p j}\, \of{ \dot h^{TT}_{r i, q }  -  N^T_{r, i q} } + \of{-\ddot h^{TT}_{k r} + 2 \, \dot N_{k,r} + 2 \, n_{, k r} } \, \of{ h^{TT}_{iq, j p} + 1/3 \, h^T_{,j p} \, \delta_{i q}} \biggr] \nonumber \\
\R^{0\ell} &=& -2 \, \eps \, \eps \, \sqof{ \dot h^{TT}_{s k, r} - N^T_{s, k r}  } \, \of{ h^{TT}_{i q, j p} +1/3 \, h^T_{,j p} \, \delta_{i q}}, \, \, \, \, \, \, \, \, \,
\eps \eps \equiv \eps^{0ijkl} \, \eps^{0 pqrs}, \nonumber \\
\R^{00} &=& -8 \, \eps \, \eps\, \sqof{ h^{TT}_{i s, l r} \, h^{TT}_{j q , k p} + 1/3 \, h^{TT}_{i s, l r} \, h^T_{,k p} \, \delta_{j q} 
+1/3\,  h^{TT}_{j q, k p} \, h^T_{,l r} \, \delta_{i s} + 1/9 \, h^T_{,l r} \, h^T_{,k p} \, \delta_{is} \, \delta_{j q}}.
\end{eqnarray}
Consider first the dynamical, $\R^{ls}$, equations. The corresponding part of the action, $\int d^5 x \, \R^{ls} \, h_{ls}$ can, upon integrating by parts, be put in the schematic form (dropping the ``TT") $\int d^5x \, \dd \dot h_{. . } [ \dd  \dot h_{. .} \,   h_{. .} + \dot h_{. .} \,  \dd  h_{. .}]$ where the $\dd$ indicate spatial derivatives and all indices contract into the implicit $\eps^{0ijkl}\,  \eps^{0pqrs}$. This is clearly far enough from ``$\dot q^2$" to make a useful kinetic term unlikely. The remaining, $\R^{0 \mu}$, equations are manifestly constraints: 
$(\R^{00}, \R^{0i})$ have (no, one)time derivatives. The obvious question is, as in the GR analysis, whether we can solve the constraints explicitly enough to exhibit the reduced action in terms of the pure ``TT" variables.  Unfortunately, these quadratic constraints are not as simple as the linear ones of GR: the ``energy--$00$" constraint is essentially (omitting spatial derivatives)
\begin{equation}
\R^{00} = \eps^{0ijkl} \, \eps^{0 pqrs}  \, R_{ilrs} \, R_{j k p q} \sim a\, (h^T)^2 + b\, h^T \, h^{TT} +c\,  (h^{TT})^2 = 0;
\end{equation}
a sum quadratic in (second derivatives of) the surviving $h_{ij}$ components. We can immediately exclude, just by indices, the mixed term: it is some complete contraction of $h^T_{,kp} \,  h^{TT}_{is,lr}$ -- but this means that at least two of the indices in $h^{TT}$ must contract internally, hence it vanishes by ``TT-ness". The first term is easily seen, upon contracting the two epsilons, to be proportional to the combination $C \equiv  [(\nabla^2\,  h^T)^2 - (h^T_{, ij})^2]$. Finally the last term is  $\sim h^{TT}_{is,lr} \, h^{TT}_{jq,kp}$, all indices necessarily contracting ``across" the two $h^{TT}$. The upshot, then, is that $C=  (h^{TT})^2$ is a time-independent equation to determine $h^T$ in terms of the $h^{TT}$; however, the combination $C$ is notoriously different from that of GR's $\nabla^2\,  h^T$. We therefore stop here, without attempting any detailed solution of this difficult constraint, one which is clearly not improved by other variable choices.  Including the Einstein term would only improve solubility by providing a linear perturbative starting-point.

\section{Einstein$+$LL system}

So far, we have analyzed the pure LL model, since the Einstein action's ADM properties need no reviewing. Physically, of course the superstring's $\alpha'$ expansion reduces, in the spin $2$ sector, to the sum of Einstein plus LL actions for the ($D=10$) metric tensor field [2]. Formally, the total action's canonical form is just the sum of the two separate ones, guaranteed by diffeomorphism invariance to maintain the ADM ``$L=p\, \dot q -N_\mu \, R^\mu$" form.  Actually, things are more complicated: $L$ must be understood as written in its original second, rather than GR's first, order form, since  the transition from the ``velocities" $K_{ij}$  of (6) to their canonical momenta becomes more complicated -- indeed, not explicitly feasible -- here\footnote{Some of these issues were previously discussed for $R+$LL in [5] using -- and simply appropriating without any credit -- the full ADM apparatus.
}. In this Section,  we will consider only R$+$LL's stability, in particular whether it has positive energy. As background, recall that any bosonic theory -- like GR -- that has a SUGRA extension and (equally essential) is non-ghost, is guaranteed to have positive energy since any SUSY model's total energy $E$ is the square of its supercharge [9]. The latter is Hermitian if the Fermionic Hilbert space metric is positive, i.e., non-ghost. But since the superstring is locally supersymmetric [10], so is every power in its $\alpha'$ expansion.  Note in this connection that the relative sign of the $R$ and LL actions is uniquely determined by the underlying string action; this means that only that one combination is guaranteed to have $E>0$. [The fields' super-transformation rules in R$+$LL need not be, and probably are not, the same as those of pure GR SUGRA.] Surprisingly, the explicit SUGRA extension of R$+$LL seems never to have been carried out explicitly; still, its existence and hence that $E>0$, is guaranteed by its superstring origin.    
Despite this guarantee, even pure GR's Hamiltonian is so far from being manifestly positive that many decades' attempts could not establish it directly. Unsurprisingly, this also turns out to be the case here, as we now illustrate, partly to display some details of the energy constraint, whose spatial integral is the energy. It has the familiar Poisson equation form 
\begin{equation}
-\nabla^2 g^T= \rho_R +\rho_{\hbox{\tiny LL}},    
\end{equation}
where the sources are from $R$ and LL respectively; the left side -- the ``energy density" -- comes (only) from $R$. We deal only with the weak field approximation below, since even its sign will prove too difficult to establish easily. Upon inserting the first order information that $g^T=0$ into the right side, one finds that the integral of $\rho_R$ is of course just the usual positive energy ``$p_{TT}^2 + q_{TT}^2$" of a free spin $2$ field; the form of $\rho_{\hbox{\tiny LL}}$ is more complicated, and we just sketch the steps: Inserting (7) into the $(00)$ component of (4), we find
\begin{equation}
\rho_{\hbox{\tiny LL}} \sim \eps^{0ijkl}\,  \eps^{0mnpq}\,  \prefive R_{ijmn} \prefive R_{klpq},\, \, \, \, \, \, \, \, \,   \prefive R_{ijmn} \equiv R_{ijmn} -(K_{in}\, K_{jm}-K_{im}\, K_{jn}),        
\end{equation}
in agreement with [5]. The purely quadratic, $R^L \, R^L$, part of (19) is nothing but the $D=4$ (linearized) Gauss-Bonnet density, which of course integrates to zero. Hence to lowest, quadratic order, the total energy of R$+$LL  has the -- positive -- value of linearized GR's $E$. While gratifying, this is unsurprising. As soon as we go beyond this trivial level, matters (unsurprisingly) become complicated.  Even pure GR's energy positivity to quartic order is not explicitly demonstrable [11]; while cubic order cannot be formally positive, by its nature. Interestingly, there is one quartic component  in the $\rho_{\hbox{\tiny LL}}$ contribution
in (19) that can be disposed of locally, namely the $K^4$ term.
Since $K$ is TT, then at each point, the tensor, and hence its powers,
lives in the $D=3$ subspace orthogonal to the propagation vector.
So the epsilon's indices cannot be saturated, and this contribution
vanishes. In any
case, the overall argument from SUSY and ghost-freedom suffices to establish that the full theory's $E>0$.

\section{Summary}
We have performed an ADM space-time decomposition of the pure LL model, to reveal a most ``non-Hamiltonian", though still diffeo-invariant, system: even the ``linearized" action is necessarily cubic. Nevertheless, we could exhibit the lower time derivative constraint structure required for any system with this invariance and the ensuing ``Bianchi" identities.  One by-product was an alternate angle on Birkhoff's theorem: only the ``TT" metric components carry time derivatives; but these vanish identically for spherically symmetric tensors. 
In the weak field limit, it was possible to express the system in terms of the metric deviation's ``TT" decomposition, though the constraints, being purely quadratic, were too difficult to solve explicitly, unlike for lowest order GR.

Separately, we considered the positive energy question for the more physical R$+$LL action.  We argued that its supersymmetrizability plus ghost-freedom imply energy positivity -- hence stability -- for the relative sign of $R$ and LL dictated by the $\alpha'$ expansion.
Although it is no less impossible to establish positivity explicitly here than in GR, we were at least able to exhibit some explicit favorable indications from the combined system's energy constraints.

\section{Acknowledgments}

The work of SD was supported in part by NSF PHY-1064302 and DOE DE-FG02-164 92ER40701 grants.

\end{document}